\documentclass[12pt,preprint]{aastex}

\begin{document}

\title{Classifying Luyten Stars Using An Optical-Infrared Reduced Proper 
Motion Diagram}

\author{Samir Salim and Andrew Gould}
\affil{Department of Astronomy, The Ohio State University,
140 W.\ 18th Ave., Columbus, OH 43210}
\email
{samir,gould@astronomy.ohio-state.edu}

\singlespace

\begin{abstract}

	We present a $V-J$ reduced proper motion (RPM) diagram for
stars in the {\it New Luyten Two-Tenths} (NLTT) catalog.  In sharp
contrast to the RPM diagram based on the original NLTT data, this 
optical-infrared RPM diagram shows distinct tracks for white dwarfs,
subdwarfs, and main-sequence stars.  It thereby permits the
identification of white-dwarf and subdwarf candidates that have a
high probability of being genuine.

\end{abstract}
\keywords{astrometry --  subdwarfs -- white dwarfs}
 
\section{Introduction
\label{sec:intro}}

	One of the principal motivations for undertaking a proper-motion
survey is to construct a reduced proper motion (RPM) diagram, in which
the RPM $(=m + 5\log \mu)$, is plotted against color.  Here $m$ is the
apparent magnitude and $\mu$ is the proper motion measured in arcsec per
year.  The RPM serves as a rough proxy for the absolute magnitude
$M = m + 5\log\pi + 5$, where $\pi$ is the parallax measured in arcsec.
Indeed, if all stars had the same transverse speed, the RPM diagram
would be identical to a conventional color-magnitude diagram (CMD) 
up to a zero-point offset.

	Hence, one can hope to roughly classify stars using a RPM diagram,
even when one has no parallax or spectroscopic information.  In particular,
subdwarfs (SDs) should be especially easy to distinguish from main-sequence 
(MS) stars, since they are several magnitudes dimmer at the same color, and 
are  generally moving several times faster.  Each of these effects tends to
move SDs several magnitudes below the MS on a RPM
diagram.  White dwarfs (WDs) should also be easily distinguished from MS
stars, since they are typically 10 magnitudes fainter at the same color,
and are generally moving at similar velocities.  Although WDs are closer
to SDs on an RPM than they are to the MS, they still should be distinguishable.

\section{NLTT RPMs: Promise and Limitations
\label{sec:nltt}}

\begin{figure}
\plotone{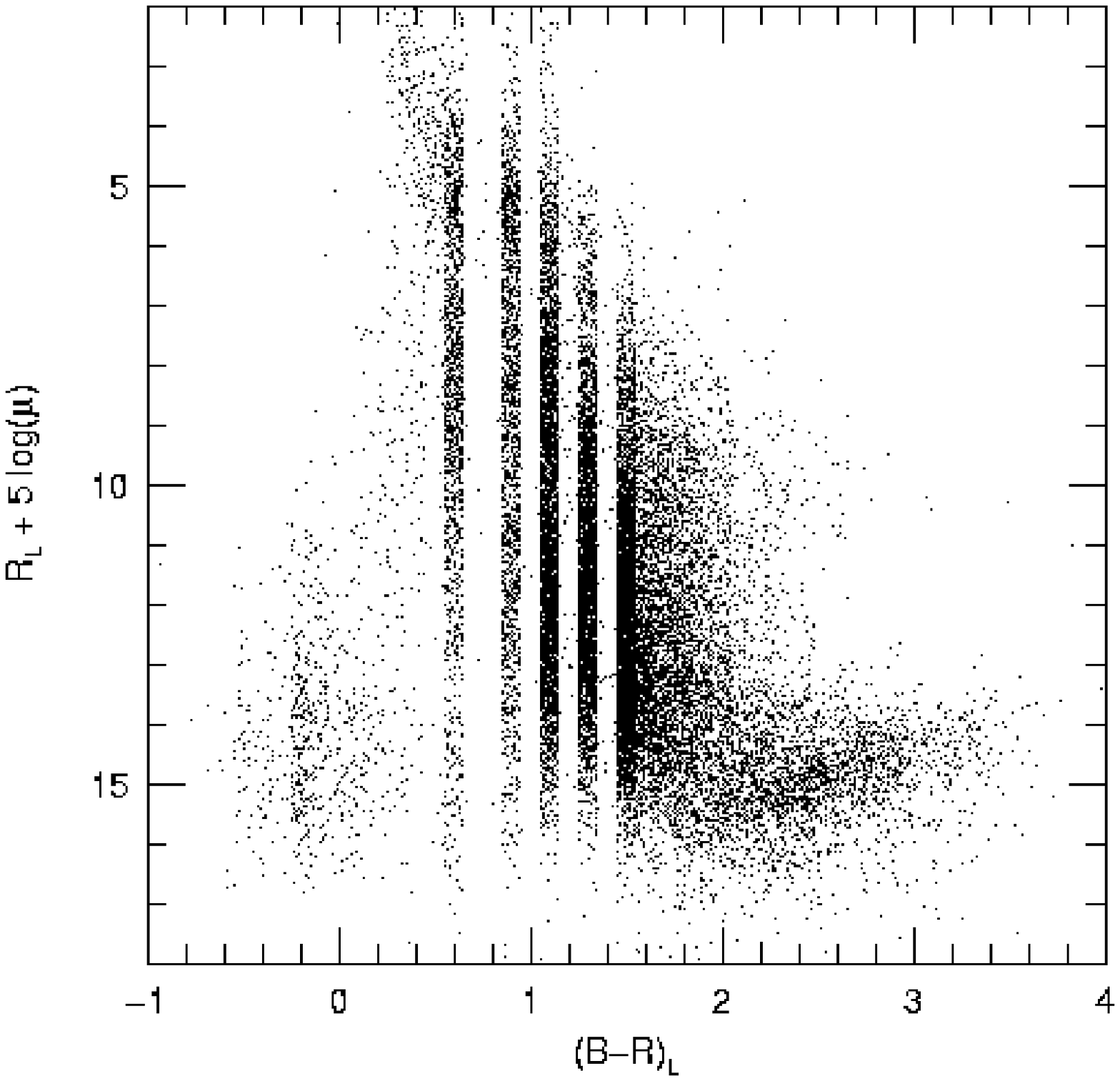}
\caption{\label{fig:rpml}
Original reduced proper motion (RPM) diagram for NLTT stars.  Both the 
proper motions,
$\mu$, and the photographic magnitudes, $B_{\rm L}$ and $R_{\rm L}$, are 
taken from NLTT.  However, since the magnitudes are originally given only to
one decimal place, the color has been randomized by 0.1 mag to show the
density of points.  The apparent stripes result from discretization of
color in the original NLTT data.
Only stars with declinations $\delta>-17^\circ$ and lying 
in areas
covered by the 2MASS release are displayed, so as to maintain consistency
with subsequent figures.
}\end{figure}

	More than two decades after its final compilation, the {\it New Luyten
Two-Tenths Catalog} (NLTT) of high proper-motion stars 
($\mu>0\farcs18\, \rm yr^{-1}$)
\citep{luy,1st}, and its better known subset {\it Luyten Half-Second Catalog}
\citep[LHS]{lhs} ($\mu>0\farcs5\, \rm yr^{-1}$) 
continue to be a vital source of
astrometric data.  They are still mined for nearby stars
\citep{reidcruz,jahr,scholz,gizreid,henry}, subdwarfs
\citep{gizreid,ryan2,ryan1}, and white dwarfs
\citep{rls,schmidt,ldgs,jones,lwd}.  NLTT is by far the largest all-sky
catalog of high-proper motion stars with more than 58,000 entries, and it
is substantially complete to $V\sim 18.5$ over most of the sky.  In addition
to positions and proper motions (PPM), NLTT contains photographic photometry in
two bands, $B_{\rm L}$ and $R_{\rm L}$, 
as well as important notes on some individual stars,
particularly binaries.  The combination of NLTT photometry and astrometry
permits the construction of a RPM diagram.  See Figure \ref{fig:rpml}.
Unfortunately, this diagram is almost devoid of features that could
be used to isolate individual populations.  There is a clump of stars off
to the lower left that one might plausibly identify with WDs.  However,
there is no obvious separation between SDs and MS stars.  Moreover, it
is common knowledge that reddish WDs are mixed in with the SDs at 
intermediate colors, so that WDs must be spectroscopically culled from 
relatively large samples drawn from the lower reaches of this diagram
(e.g., \citealt{ldgs}).
There are three reasons for this mixing of populations.  First, photographic
$B$ and $R$ do not provide a very broad color baseline.  Second, 
photographic photometry has intrinsically large errors.  Third, NLTT
does not reach even the relatively limited precision that is 
in principle achievable 
with photography.  Given the steepness of the color-magnitude relations
of individual populations, the short color baseline and large errors
combine to seriously smear out the diagram.

\section{New Photometry and Astrometry for NLTT Stars
\label{sec:rpm}}

	We are in the process of obtaining improved photometry and
astrometry for the great majority of NLTT stars.  Here we give a
very brief summary of our procedure. Details are given in \citet{gs}
and \citet{sg}.   For bright stars, we find counterparts to NLTT stars in 
three PPM catalogs: Hipparcos \citep{hip}, Tycho-2 \citep{t2}, and Starnet 
\citep{starnet}.
Using this improved astrometry, we easily identify the 2MASS \citep{2mass}
second incremental release 
counterparts of these stars and so obtain infrared (IR) photometry.
Hipparcos and Tycho-2 have generally excellent optical photometry.  For 
faint stars, we first identify candidate counterparts from
USNO-A \citep{usnoa1,usnoa2} around
the (circa 1950 epoch) positions given by NLTT, and then ask if there are any
2MASS counterparts around the (circa 2000 epoch) positions
predicted by the USNO-A positions and the NLTT proper motion.  We use
several criteria to ascertain the validity of the NLTT/USNO-A/2MASS match.
Thus, we obtain
infrared photometry from 2MASS, optical photometry from USNO-A, and
proper motions with typical accuracies of $5\,\rm mas\,yr^{-1}$ from
the difference of positions in the two catalogs.  The former allows us to
construct the optical-IR color index $V-J$.  The latter represents a
roughly 5-fold improvement relative to the NLTT proper-motion errors
\citep{sg}.

\begin{figure}
\plotone{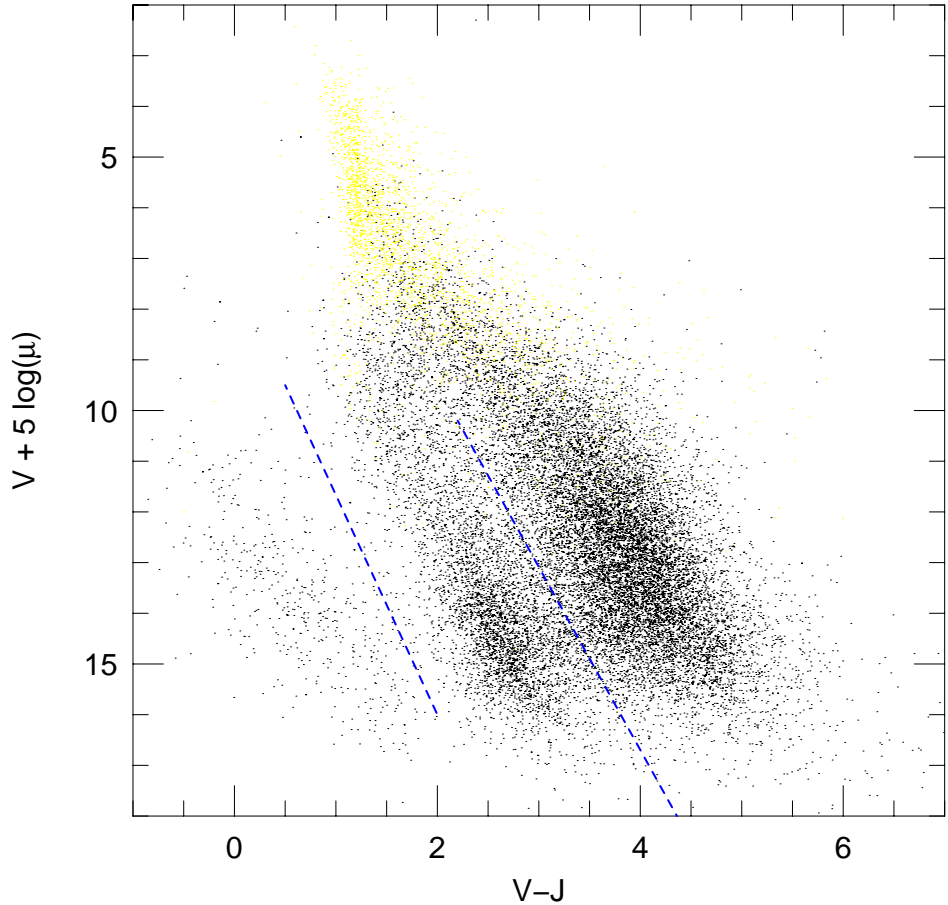}
\caption{\label{fig:rpmv}
Optical-infrared RPM diagram for NLTT stars with $\delta>-17^\circ$,
and having 2MASS and Hipparcos counterparts ({\it yellow}),
as identified by \citet{gs}, or 2MASS and USNO-A counterparts ({\it black}),
as identified by \citet{sg}.  For RPM $V+5\log\mu\ga 9$, subdwarfs (SDs) and 
main-sequence (MS) stars clearly lie on different tracks.  White dwarfs
(WDs) are also clearly separated from SDs and MS stars.  In the subsequent
figure, the NLTT stars with 2MASS/USNO-A counterparts will be identified
as MS (red), SD (green), or WD (cyan), depending on whether they lie
to the right, between, or to the left of the two dashed lines shown in 
this figure.
}\end{figure}

	Figure \ref{fig:rpmv} shows the resulting RPM restricted to
$\delta>-17^\circ$, for which the data of \citet{gs} and \citet{sg} are
relatively homogeneous.  For $V$ band we adopt \citep{nearbylens} $V = R_{\rm
U} + 0.23 + 0.32(B-R)_{\rm U}$, where $B_{\rm U}$ and $R_{\rm U}$ are the
photographic magnitudes from USNO-A.  Note that the SDs and MS stars are
clearly separated into two tracks, at least for $V+5\log\mu\ga 9$.  The WDs
are also clearly separated from the other stars.  Going towards the bright end, the SD track becomes
vertical at $V+5\log\mu\sim 10$ and then turns to the right at $V+5\log\mu\sim
9$.  For a star having a transverse speed of $v_\perp\sim 250\,\rm km\,
s^{-1}$, this latter value corresponds to $M_V\sim 5$, i.e., roughly a mag
below the subdwarf turnoff.  In any event, the bright end is dominated by
Hipparcos counterparts of NLTT stars (shown in yellow).  Hipparcos stars
generally have excellent parallaxes, so for them RPMs are superfluous.  We
therefore remove these stars from further consideration.

	Using superior color-color plots of NLTT stars based on Sloan Digital
Sky Survey (SDSS, \citealt{sdss}) photometry, we have checked that the scheme
shown in Figure \ref{fig:rpmv} indeed properly classifies NLTT stars that lie
in the SDSS Early Data Release \citep{stoughton} area.  In a future paper, we
will show how SDSS/USNO-A data can be combined to classify stars at much lower
proper motions than the threshold of NLTT.

	Once stars are classified using the RPM diagram, it is possible to
estimate their distances photometrically.  Using this technique, in Table
\ref{table:wds} we have tentatively identified 23 WD
candidates\footnote{Here we extend our sample to $\delta>-33\degr$} that lie within
20 pc and are not listed in the online edition of the Gliese Catalog of Nearby
Stars\footnote{http://www.ari.uni-heidelberg.de/aricns}. These WD
candidates are also not listed in \citet{mccook} catalog of spectroscopically
confirmed WDs. We determine photometric distances from $J$, $V-J$, and a CMR
calibrated using \citet{bergeron} data. Stars with low values of $\Delta(V-J)$
lie close to the SD/WD discriminator line shown in Figure \ref{fig:rpmv}, and
could therefore be extreme SDs--stars interesting in their own
right. Spectroscopy will be ultimately required to confirm the nature of these
objects.  If confirmed, these WDs would represent a significant addition to
109 WDs known to be closer than 20 pc \citep{hos}. In fact, since for 10 of
our candidate WDs we calculate $d<13$ pc, they could raise the local density
derived by \citet{hos} using a 13 pc sample (which they believe is complete)
by as much as 20\% (but note that we do not cover the entire sky).

\section{Comparison with NLTT
\label{sec:compare}}

\begin{figure}
\epsscale{0.65}
\plotone{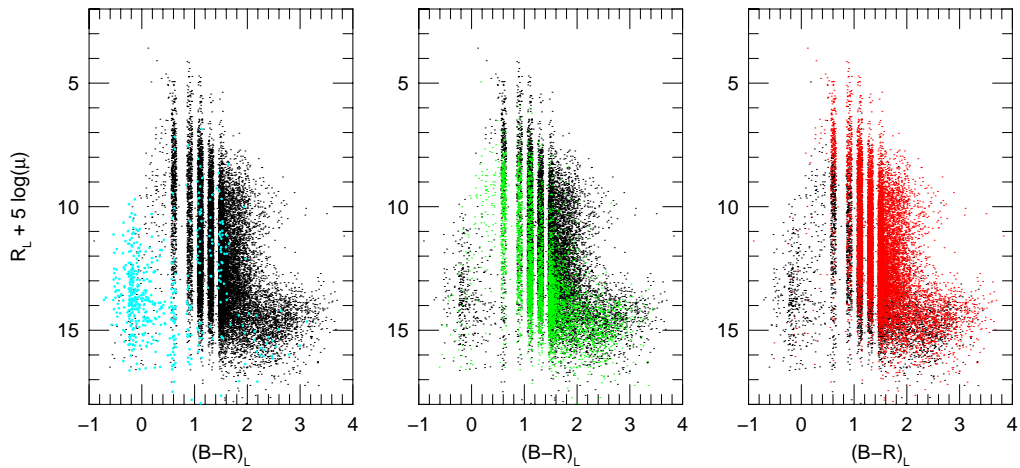}
\caption{\label{fig:rpmlall}
Left panel: Original NLTT RPM diagram (see Fig.\ \ref{fig:rpml}) but with the stars
identified in Fig.\ \ref{fig:rpmv} as WDs shown as large cyan dots.  While the region to the lower left is strongly dominated by WDs, many
WDs lie further to the red where they are heavily contaminated by other stars.
SDs are shown as small green dots in the middle panel, and MS stars are shown
as small red dots in right panel. Comparison of SD and MS distribution shows
that they strongly overlap in NLTT RPM. In all cases the other types of stars
with 2MASS/USNO-A counterparts are shown as smaller black dots.
}\end{figure}

	Panels in Figure \ref{fig:rpmlall} show respectively
where the stars classified as WDs, SDs, and MS stars in Figure \ref{fig:rpmv},
lie in the original NLTT RPM.  The points representing the WDs are enlarged
for emphasis.  Note that while the region to the lower left is indeed
dominated by WDs, the red WDs are sprinkled among a much higher density
of other faint red stars. From the distribution of SDs and MS stars
it is clear that while the SDs do tend on average to be bluer and fainter 
than the MS stars, the two populations are severely mixed in the original
NLTT RPM diagram.

\section{Conclusions}
\label{sec:conclusion}

	We have presented an optical/IR RPM diagram for NLTT stars
based on photometry and astrometry obtained from USNO-A and 2MASS.
WDs, SDs, and MS stars are well separated in this diagram, at least
for stars with RPMs $V+5\log\mu\ga 9$.  This should provide a powerful
tool for choosing WD and SD candidates from among the 58,845 NLTT
stars for a variety of studies.

\acknowledgments We acknowledge the use of CD-ROM versions of USNO-A1.0 and
A2.0 catalogs provided to us by D.\ Monet.  This publication makes use of
catalogs from the Astronomical Data Center at NASA Goddard Space Flight
Center, VizieR and SIMBAD databases operated at CDS, Strasbourg, France, and
data products from the Two Micron All Sky Survey, which is a joint project of
the University of Massachusetts and the IPAC/Caltech, funded by the NASA and
the NSF.  We thank the referee for useful suggestions. This work was supported
by JPL contract 1226901.


\clearpage
\begin{deluxetable}{r r r r r r r r r r r r l}
\tabletypesize{\scriptsize}
\tablecaption{New candidate white dwarfs closer than 20 pc \label{table:wds}}
\tablewidth{0pt}
\tablehead{
\colhead{NLTT} &
\multicolumn{3}{c}{R.A.}   &
\multicolumn{3}{c}{Decl.}   &
\colhead{$V$} &
\colhead{$B-V$}  &
\colhead{$V-J$} &
\colhead{$\Delta(V-J)$} &
\colhead{$d$ (pc)} &
\colhead{Notes}  \\
}
\startdata
21351 & 09 & 16 & 35.85 & +19 & 55 & 17.5 & 12.8  &   1.5 &  0.58 &  0.0 &    6.6 & \tablenotemark{a}  \\
52890 & 22 & 05 & 33.03 & +19 & 51 & 24.7 & 13.4  &   1.0 &  0.78 &  0.1 &    7.3 & \tablenotemark{b} \\
44850 & 17 & 29 & 17.38 & $-$30 & 48 & 36.8 & 11.8  & \nodata & 0.02 & 0.4 &    7.5 & \tablenotemark{c} \\
24703 & 10 & 34 & 02.98 & +12 & 10 & 15.1 & 13.4  &   1.1 &  0.70 &  0.0 &    7.7 & \tablenotemark{b} \\
47807 & 19 & 27 & 21.31 & $-$31 & 53 & 06.1 & 13.2  &   0.9 &  0.51 &  0.1 &    8.4 & \\
57038 & 23 & 29 & 43.45 & $-$09 & 29 & 55.8 & 12.6  &   0.9 &  0.13 &  0.2 &    9.5 & \tablenotemark{b} \\
26398 & 11 & 08 & 14.50 & $-$14 & 02 & 47.1 & 14.2  &   0.8 &  0.73 &  0.3 &   11.0 & \tablenotemark{b} \\
32147 & 12 & 52 & 10.43 & +18 & 33 & 09.4 & 14.1  &   1.3 &  0.63 &  0.3 &   11.7 & \\
 4828 & 01 & 26 & 48.89 & $-$26 & 33 & 55.7 & 14.6  &   0.9 &  0.89 &  0.2 &   11.8 & \\
15314 & 05 & 34 & 20.18 & $-$32 & 28 & 51.3 & 11.6  &   0.8 & $-$0.41 &  1.0 &   11.8 & \tablenotemark{b} \\
11551 & 03 & 41 & 21.11 & +42 & 52 & 33.9 & 15.8  &   1.2 &  1.40 &  0.1 &   13.9 & \tablenotemark{b} \\
56805 & 23 & 25 & 19.86 & +14 & 03 & 39.4 & 15.8  &   1.2 &  1.37 &  0.1 &   14.2 & \\
 8435 & 02 & 35 & 21.79 & $-$24 & 00 & 47.0 & 15.8  &   0.6 &  1.32 &  0.5 &   14.3 & \tablenotemark{d} \\
23235 & 10 & 01 & 40.00 & $-$24 & 04 & 41.2 & 15.8  &   1.5 &  1.18 &  0.3 &   15.7 & \tablenotemark{e} \\
19138 & 08 & 14 & 11.15 & +48 & 45 & 29.8 & 15.1  &   0.8 &  0.76 &  0.5 &   16.1 & \\
47373 & 19 & 04 & 50.62 & $-$03 & 43 & 07.0 & 14.6  &   1.9 &  0.49 &  0.8 &   16.8 & \\
46292 & 18 & 20 & 02.58 & $-$27 & 45 & 50.5 & 14.6  &   0.4 &  0.46 &  0.8 &   17.3 & \\
 8581 & 02 & 39 & 19.68 & +26 & 09 & 57.5 & 16.3  &   0.8 &  1.40 &  0.2 &   17.4 & \\
34826 & 13 & 40 & 00.80 & +13 & 46 & 51.9 & 16.3  &   1.9 &  1.36 &  0.1 &   17.8 & \\
  529 & 00 & 11 & 22.45 & +42 & 40 & 40.8 & 15.2  &   0.5 &  0.69 &  0.5 &   18.4 & \tablenotemark{f} \\
53177 & 22 & 12 & 17.96 & $-$14 & 29 & 46.1 & 15.0  &   0.5 &  0.51 &  0.8 &   19.8 & \tablenotemark{g} \\
31748 & 12 & 44 & 52.66 & $-$10 & 51 & 08.9 & 14.4  &   0.5 &  0.24 &  1.1 &   19.9 & \tablenotemark{h} \\
40607 & 15 & 35 & 05.81 & +12 & 47 & 45.2 & 15.9  &   0.8 &  0.95 &  0.4 &   20.0 & \tablenotemark{i}
 \\
\enddata
\tablecomments{NLTT numbers follow the record number in the electronic
version of NLTT (ADC/CDS catalog I/98A). Positions are given for epoch and
equinox 2000. $V$ and $B-V$ are calibrated from USNO photographic magnitudes,
except for NLTT 44850 (GSC red magnitude). $\Delta (V-J)$ gives horizontal
separation from the SD/WD discriminator line. Distances are photometric, based
on \citet{bergeron} data.}

\tablenotetext{a}{USNO photometry blended with NLTT 21352.}
\tablenotetext{b}{NLTT magnitudes fainter.}
\tablenotetext{c}{2MASS photometry possibly affected by crowding.}
\tablenotetext{d}{LHS 1421.}
\tablenotetext{e}{USNO photometry blended with NLTT 23234.}
\tablenotetext{f}{\citet{giclas} : Suspected WD (GD 5).}
\tablenotetext{g}{\citet{beers} : $V=15.09$, $B-V=0.23$, `composite' spectrum.}
\tablenotetext{h}{\citet{green} : Spectral type sdB (PG 1242--106).}
\tablenotetext{i}{\citet{eggen} : $V=15.07$, $B-V=0.24$ (WD 1532+12).}

\end{deluxetable}

\end{document}